\documentclass[a4paper,twocolumn,11pt,accepted=2026-11-23]{quantumarticle}
\pdfoutput=1
\usepackage[utf8]{inputenc}
\usepackage[english]{babel}
\usepackage[T1]{fontenc}
\usepackage{amsmath}
\usepackage{hyperref}
\usepackage[numbers]{natbib}
\usepackage{amssymb}

\begin{document}

\title{Fast quantum-state transfer in interacting topological Su-Schrieffer-Heeger chains}

\author{François Impens}
\email{impens@if.ufrj.br}
\affiliation{Instituto de F\'{\i}sica, Universidade Federal do Rio de Janeiro, 21941-972 Rio de Janeiro, RJ, Brazil}
\orcid{0000-0002-4949-9717}
\author{David Guéry-Odelin}
\email{dgo@irsamc.ups-tlse.fr}
\affiliation{Laboratoire Collisions Agrégats Réactivité, UMR 5589, FERMI, Université de Toulouse, CNRS, 118 Route de Narbonne, 31062 Toulouse CEDEX 09, France.}
\orcid{0000-0003-4869-1341}

\maketitle


\begin{abstract}
 Shortcuts to adiabaticity based on next-nearest-neighbor hoppings enable fast single-particle state transfer via topological edge states in Su-Schrieffer-Heeger chains. We show that making these hoppings phase-tunable extends the shortcut to interacting many-particle systems. In the mean-field regime, this yields an exact nonlinear shortcut: one hopping quadrature keeps the state on the instantaneous dark-state trajectory, while the orthogonal quadrature cancels the interaction-induced self-phase modulation. The resulting protocol is nonperturbative in the mean-field interaction strength. When applied to the full Bose-Hubbard dynamics, the mean-field shortcut remains beneficial but saturates below unit fidelity, exposing genuinely many-body corrections beyond the product-state picture. We then optimize the transfer directly in the many-body Hilbert space and find that complex, phase-tunable next-nearest-neighbor hoppings recover near-perfect fidelity. Our results show that hopping phases are not merely a technical refinement, but a key control resource for fast and high-fidelity transport in interacting topological systems.
\end{abstract}

\section{INTRODUCTION}

Beyond their importance in fundamental physics, topological quantum systems
provide a promising platform for robust quantum-state manipulation and transport~\cite{Kitaev01, TopologicalNetwork17},
owing to the resilience of edge states against disorder and imperfections. One
of the simplest paradigmatic models exhibiting such states is the
Su-Schrieffer-Heeger (SSH) chain~\cite{SSH79}, experimentally implemented or emulated in different systems~\cite{StJean17,Science2018,AtomicPlatform1,Gadway19,Ronco26}. In this context,
quantum-state transfer (QST) between distant edge or interface modes has attracted considerable
attention and motivated the development of resilient adiabatic transfer protocols~\cite{Longhi19a,Longhi19b,Li2025}. Since long operation times increase the susceptibility to decoherence
and losses, accelerating such transfers while preserving high fidelity is highly
desirable.

Next-nearest-neighbor~(NNN) assisted shortcuts to adiabaticity provide an elegant way to accelerate this transfer in the single-particle regime~\cite{Dangelis20}. Since then, shortcut protocols and accelerated-transfer methods have extended this idea to several topological platforms in the single-excitation context, ranging from optomechanical systems to coupled emitter arrays~\cite{Palaiodimopoulos2021,Qi2021Router,Qi2021BeamSplitter,Baum2022,Huang2022,Zurita2023,Guo2024,Han2024,Tian2024,Xu2024Spectrum,Romero2024,Yao2024,Kanda2025}.  Related acceleration strategies have been explored for classical topological pumping~\cite{Nori2020,Takahashi2020,Impens2025}. Beyond the single-excitation setting, recent works have addressed Gaussian-state transfer~\cite{Hao2025,Fang2026}, and nonlinear pumping~\cite{You2025} in topological systems. 
Enhanced two-excitation quantum-state transfer has also
been demonstrated experimentally in superconducting-qubit
networks, using optimized couplings to circumvent
quantum-chaotic dynamics~\cite{Xiang2024}.

So far, however, shortcuts to adiabaticity and other fast quantum-state-transfer methods in topological systems have remained restricted to the linear, non-interacting regime. Going beyond this limitation is not straightforward: interactions do not merely perturb the linear transfer problem, but can qualitatively modify the structure of one-dimensional quantum systems and have been shown, for instance, to affect their topology in both Hermitian~\cite{Ezawa2021} and non-Hermitian settings~\cite{Xi21,Kawabata22,Arouca26}. Bypassing the linear setting is important because many experimentally relevant
platforms, including ultracold atoms in optical lattices~\cite{Jaksch98},
long-range interacting quantum simulators~\cite{Defenu23}, and nonlinear
photonic systems~\cite{Smirnova20}, naturally operate in regimes where
interactions cannot be neglected. This raises two central questions: can fast quantum-state transfer methods be successfully extended to nonlinear topological systems, and, if so, how do interactions modify the control resources required for high-fidelity state transfer in topological chains?

A natural first step toward answering these questions is provided by a mean-field description. In the linear SSH chain, suitably engineered NNN hoppings on the relevant sublattice can promote the instantaneous edge-state profile from an adiabatic solution to an exact finite-time trajectory. We show that the edge-state density profile underlying this shortcut survives in the nonlinear regime: interactions do not deform the instantaneous edge-state populations. Their effect is instead to generate site-dependent nonlinear phase shifts, which cannot be compensated by the usual NNN quadrature of the linear shortcut. We derive an exact analytical extension of the NNN-assisted shortcut method in which one quadrature cancels nonadiabatic transitions, while an additional orthogonal quadrature compensates the interaction-induced phase shifts.

The paper is organized as follows. In Sec.~\ref{sec:linearshortcut}, we first recall the interaction-free NNN-assisted shortcut construction in SSH chains, which provides the reference protocol and the starting point for the interacting extensions. In Sec.~\ref{sec:meanfield}, we show how this approach can be generalized exactly in the nonlinear mean-field regime by introducing a second, phase-tunable quadrature of the NNN hoppings; we then benchmark the resulting protocol both at the mean-field level and under the full Bose-Hubbard dynamics. In Sec.~\ref{sec:PMPmanybody}, we move beyond mean-field trajectory engineering and use Pontryagin-based optimization directly in the many-body Hilbert space. This allows us to identify phase-tunable NNN hoppings as an effective control resource for restoring high-fidelity transfer in the genuinely interacting regime.

\section{Interaction-free NNN-assisted shortcuts in SSH chains}
\label{sec:linearshortcut}

\subsection{Quantum state transfer in SSH chains}

\begin{figure}[htbp]
    \centering
    \includegraphics[width=8.7cm]{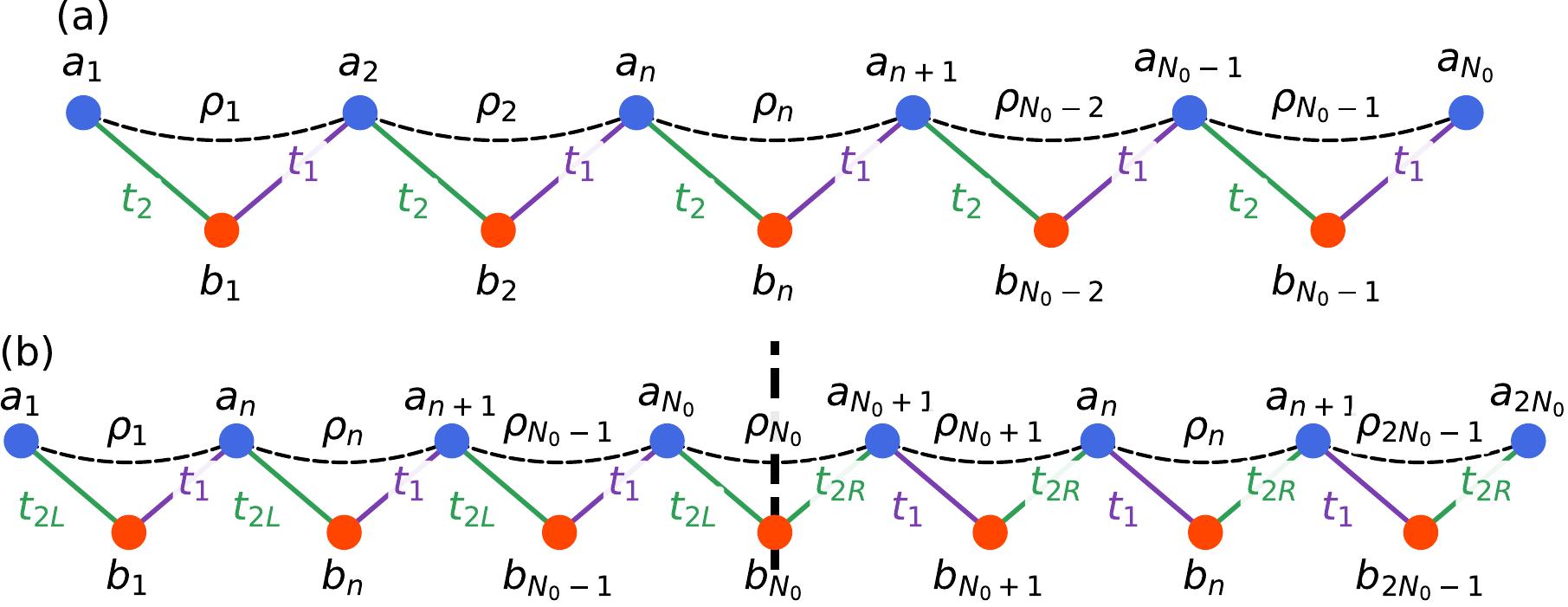}
    \caption{ (a) Simple SSH chain with $N=2 N_0-1$ sites with $t_1$~(purple) and $t_2$~(green) NN hoppings.
    (b) SSH chain with a topological interface at the central $B_{N_0}$ site, with $N=4N_0-1$ sites. In both SSH geometries, $N_\rho$ tunable NNN couplings $\rho_n(t)$ connect adjacent sites of the $A$-sublattice, with $N_\rho=N_0-1$ in the simple SSH chain and $N_\rho=2N_0-1$ in the topological-interface geometry.  We first consider single-particle SSH chains with real-valued NNN couplings, and then extend the discussion to the many-particle case with phase-tunable complex-valued NNN couplings $\rho_n(t)$.}
    \label{fig:SSHscheme}
\end{figure}

The Su--Schrieffer--Heeger (SSH) chain~\cite{SSH79}
is a paradigmatic one-dimensional lattice model supporting
topologically protected edge states. The two SSH geometries considered throughout the paper are shown in Fig.~\ref{fig:SSHscheme}(a,b): a simple finite SSH chain and an SSH chain containing a topological interface. 

We first consider a simple SSH chain with $N=2N_0-1$ sites and time-dependent nearest-neighbor hoppings $t_1(t)$ and $t_2(t)$,
\begin{equation}
\hat H_0(t)=
\sum_{n=1}^{N_0-1}
\left[
t_2(t)|B_n\rangle\langle A_n|
+
t_1(t)|A_{n+1}\rangle\langle B_n|
\right]
+\mathrm{H.c.}
\label{eq:singleSSHHamiltonian}
\end{equation}
The sites $A_n$ and $B_n$ define the two sublattices of the SSH chain, and we denote by $|A_n\rangle$ and $|B_n\rangle$ the associated localized single-particle states. The Hamiltonian $\hat{H}_0$~\eqref{eq:singleSSHHamiltonian} possesses chiral symmetry, $\{ \hat \Gamma, \hat H_0 \}=0$, with the chiral operator $\hat \Gamma= \sum_{n=1}^{N_0} | A_n \rangle \langle A_n | -  \sum_{n=1}^{N_0-1} | B_n \rangle \langle B_n |.$ For an SSH chain of odd parity, thus presenting an imbalance between the two sublattices, this symmetry results in a zero-energy mode entirely supported on the $A$ sublattice, given by 
\begin{eqnarray}
| \phi_0(t) \rangle & = & \mathcal{N}(t) \sum_{n=1}^{N_0} \: \left( \frac {-t_2(t)} {t_1(t)} \right)^{n-1} | A_n \rangle 
\label{eq:zero_edge_mode}
\end{eqnarray} 
for $t_1(t)\neq 0$, where $\mathcal{N}(t)$ is a normalization factor.

For $|t_2(0)/t_1(0)|\ll 1$, the zero-energy mode~\eqref{eq:zero_edge_mode} is
initially localized near the left boundary of the chain. By slowly increasing the ratio $|\epsilon(t)|=|t_2(t)/t_1(t)|$ from $|\epsilon(0)|\ll 1$ up to a final value such that  $|\epsilon(T)|\gg 1$, this mode is adiabatically displaced from the left  to the right boundary. Thus, a single-particle
state initially localized at the left end,
$|\psi(0)\rangle=|A_1\rangle$, can be transported
with high fidelity to the opposite end,
$|\psi(T)\rangle\simeq |A_{N_0}\rangle$,
provided the evolution is sufficiently slow
so that nonadiabatic transitions remain marginal.

\subsection{Shortcut transfer in a single-particle SSH chain}

We first recall the interaction-free shortcuts in the SSH context~\cite{Dangelis20}, which will serve as a basis to build the interacting many-body protocols developed below.

The shortcut-to-adiabaticity method consists in using additional control resources, namely  tunable NNN hoppings connecting neighboring sites of the (A) sublattice~(see Fig.~\ref{fig:SSHscheme}a), in order to cancel non-adiabatic transitions out of the instantaneous
zero-energy mode of the reference SSH Hamiltonian.
Although the NNN hoppings break the chiral symmetry
of this Hamiltonian, they allow its instantaneous
zero-mode trajectory to become an exact solution
of the finite-time Schr\"odinger dynamics. The corresponding NNN Hamiltonian reads:
\begin{equation}
\hat H_{\rm NNN}^{(1)}(t) =
\sum_{n=1}^{N_\rho}
i\rho_n(t)|A_{n+1}\rangle\langle A_n|
+\mathrm{H.c.}
\label{eq:CDHamiltonian}
\end{equation}
where $N_\rho$ is the number of independent real-valued NNN coupling amplitudes $\rho_n(t)$. For the simple SSH geometry, $N_\rho=N_0-1$. Writing the single-particle state as $ |\psi(t)\rangle=\sum_{n=1}^{N_A}a_n(t)|A_n\rangle+\sum_{n=1}^{N_B}b_n(t)|B_n\rangle $, with $N_A=N_0$ and $N_B=N_0-1$ for the simple SSH chain, the  Schr\"odinger equation yields the linear dynamical system~(hereafter, we set $\hbar=1$)
\begin{eqnarray}
i \dot a_n \! \! \!
&=& \! \!
t_2 b_n + t_1 b_{n-1} \!
- \!i\rho_n \, a_{n+1}
+ \! i \rho_{n-1} \, a_{n-1} \!, \qquad  
\label{eq:lin_an}
\\[1mm]
i \dot b_n \! \! \!
&=& \! \!
t_2 a_n + t_1 a_{n+1} .
\label{eq:lin_bn}
\end{eqnarray}
To enforce exact evolution along the instantaneous
zero-energy mode $|\phi_0(t)\rangle$, we impose
$b_n(t)=0$ at all times. The $B$-sublattice equations~\eqref{eq:lin_bn}
then reduce to the dark-state condition
\begin{equation}
\label{eq:simple_dark_state_condition}
t_2 a_n +t_1 a_{n+1} =0,
\qquad 1\le n\le N_0-1 \,,
\end{equation}
from which we infer the amplitudes $a_n(t)=\mathcal N(t)\epsilon(t)^{n-1}$
with $\epsilon(t)=-t_2(t) / t_1(t)$. One thus retrieves the zero-energy mode Eq.\eqref{eq:zero_edge_mode}.

The remaining $A$-sublattice equations~\eqref{eq:lin_an} determine the NNN couplings as
\begin{equation}
\dot a_n(t)
=
\rho_{n-1}(t)a_{n-1}(t)
-
\rho_n(t)a_{n+1}(t),
\label{eq:linear_inverse_engineering_simple}
\end{equation}
with $1\le n\le N_0$ with the boundary convention $\rho_0=\rho_{N_0}=0$. In practice, the first equation determines
$\rho_1(t)=-\dot a_1(t)/a_2(t)$, while the subsequent
equations determine the remaining NNN couplings $\rho_n(t)$
recursively. The obtained NNN couplings are real-valued, and the last equation is automatically
satisfied by the normalization of the prescribed
zero-mode trajectory~\cite{Dangelis20}. These couplings guarantee that the instantaneous zero mode $|\phi_0(t)\rangle$~\eqref{eq:zero_edge_mode} becomes an exact solution of the driven single-particle dynamics. This approach provides the interaction-free shortcut benchmark used throughout the paper.



\subsection{Shortcut transfer across a topological interface}

The same inverse-engineering strategy also applies to an SSH chain containing a topological interface. This geometry consists of two SSH segments with
opposite dimerization patterns, connected at
the central $B_{N_0}$ site~(see Fig.~\ref{fig:SSHscheme}b).

Specifically,  consider a chain of $N=4N_0-1$ sites driven by the single-excitation Hamiltonian

 \begin{align}
& \hat H_1(t)
  \! = \! 
t_{2L}(t)\sum_{n=1}^{N_0}
|B_n\rangle\langle A_n|
+
t_1(t)\sum_{n=1}^{N_0-1}
|A_{n+1}\rangle\langle B_n|
\nonumber\\
& \!  +
t_{2R}(t)  \!   \! \! \sum_{n=N_0}^{2N_0-1}  \! \! 
|A_{n+1}\rangle\langle B_n|
+
t_1(t) \! \! \! \! \sum_{n=N_0+1}^{2N_0-1}
 \! \!  |A_n\rangle\langle B_n| \!
 + \!  \mathrm{H.c.} \label{eq:interface_hamiltonian}
\end{align}
We follow the approach of Ref.~\cite{Dangelis20} to design the shortcut protocol. For the topological-interface SSH chain, the $B_n$
sites are coupled to their left and right $A$ neighbors
through the hoppings $t_{2L}$ and $t_1$ on the left
segment, and through $t_1$ and $t_{2R}$ on the right segment,
respectively. At the central $B_{N_0}$ site, these
hoppings are $t_{2L}$ and $t_{2R}$.
Requiring the $B$-sublattice amplitudes to vanish
therefore leads to the following piecewise dark-state conditions:
\(t_{2L}a_n+t_1a_{n+1}=0\) on the left segment,
\(t_1a_n+t_{2R}a_{n+1}=0\) on the right segment, and $t_{2L}a_{N_0}+t_{2R}a_{N_0+1}=0$
at the interface. This leads us to introduce $\epsilon_L(t)=-t_{2L}(t)/t_1(t)$, $\epsilon_R(t)=-t_{2R}(t)/t_1(t)$. The amplitudes

{\begin{equation}
\frac {a_n(t)} {\mathcal N_I(t)}=
\begin{cases}
\epsilon_L(t)^{n-1}\epsilon_R(t)^{N_0}, \! \!
& \! \! 1\le n\le N_0,\\
-\epsilon_L(t)^{N_0}\epsilon_R(t)^{2N_0-n}, \! \!
& \! \! N_0<n\le 2N_0.
\end{cases}
\label{eq:interfacemode}
\end{equation}}
define a zero-mode supported entirely on the $A$ sublattice. By construction, this zero-mode satisfies the dark-state condition on the left and right sides separately, as well as the interface condition.

  

The transfer across the interface is obtained by
keeping $t_1(t)=t_0$ constant and prescribing
time-dependent hoppings $t_{2L}(t)$ and $t_{2R}(t)$
symmetric with respect to the midpoint $t=T/2$,
$t_{2L}(t)=t_{2R}(T-t)$.
The corresponding ratios satisfy
$|\epsilon_L(0)|\ll 1$ and
$|\epsilon_R(T)|\ll 1$, while
$|\epsilon_{L,R}(t)|<1$ throughout the protocol. 
For the considered SSH geometry, this last condition
ensures that the two segments remain in distinct
topological phases, thus preserving the interface
throughout the protocol.

As follows from Eq.~\eqref{eq:interfacemode},
the boundary amplitudes are
$a_1(t)=\mathcal N_I(t)\epsilon_R(t)^{N_0}$ and
$a_{2N_0}(t)=-\mathcal N_I(t) \epsilon_L(t)^{N_0}$, so that $|a_{2N_0}(t)/a_1(t)|=|\epsilon_L(t)/\epsilon_R(t)|^{N_0}$.
The prescribed variation of the NN hoppings
therefore shifts the zero-mode localization
from the left boundary to the right boundary,
through a delocalized state
near the midpoint of the protocol when $|\epsilon_L(T/2)|=|\epsilon_R(T/2)|\simeq 1$. In the adiabatic limit, this produces
quantum-state transfer across the interface.
At finite transfer times, however, additional NNN
couplings are required to preserve the
instantaneous zero-mode trajectory exactly.  In contrast to the piece-wise dark-state condition, the reverse-engineering 
condition~\eqref{eq:linear_inverse_engineering_simple} remains unchanged by the presence of an interface.


Adding the NNN Hamiltonian~\eqref{eq:CDHamiltonian} with \(N_\rho=2N_0-1\),
the interface ansatz~\eqref{eq:interfacemode} becomes an exact finite-time
solution if the NNN couplings
follow the relations~\eqref{eq:linear_inverse_engineering_simple} for $1 \leq n \leq 2 N_0$ with the boundary convention $\rho_0=\rho_{2N_0}=0$. The $2N_0-1$
NNN couplings $\rho_n(t)$ are determined either by direct
recurrence~\cite{Dangelis20} or by solving the
corresponding linear system at each time. For the midpoint-symmetric NN hopping profiles considered
here, the NNN couplings fulfill $\rho_{N_\rho+1-n}(t)=\rho_n(T-t)$. Thus, this approach enables arbitrarily high-fidelity finite-time transfer across the
interface within the single-excitation sector, at the cost of independently
controlling \(2 N_0-1\) NNN couplings whose amplitudes scale as \(1/T\).

\section{Nonlinear shortcut for quantum-state transfer in an SSH chain with mean-field interactions}
\label{sec:meanfield}

We now extend the discussion to interacting bosonic SSH chains. The dynamics is described by the time-dependent Bose-Hubbard Hamiltonian~\cite{Jaksch98} whose components, for the simple SSH chain, read
\begin{align}
& \hat H_{\rm 0, SSH}(t)  
  \! \! = \!  \! \! 
\sum_{n=1}^{N_0-1} \!
\left[
t_2(t)\,\hat b_n^\dagger \hat a_n
+
t_1(t)\,\hat a_{n+1}^\dagger \hat b_n
 +
{\rm H.c.} \right],  \\
& \hat H_{\rm int}
\! \! 
  =  \! \! 
\frac{U_A}{2}\sum_{n=1}^{N_A}
\hat a_n^\dagger\hat a_n^\dagger\hat a_n\hat a_n
+
\frac{U_B}{2}\sum_{n=1}^{N_B}
\hat b_n^\dagger\hat b_n^\dagger\hat b_n\hat b_n . 
\end{align}
For convenience, in equations applying to both the simple and topological-interface SSH geometries,
we denote by $N_A$ and $N_B$ the numbers of sites on the respective sublattices. The number of NNN couplings is $N_\rho=N_A-1,$ independently of the chosen geometry.

For the SSH chain with topological interface, the SSH contribution $\hat H_{\rm 1, SSH}(t)$ is obtained by second-quantizing $\hat{H}_1$ instead of $\hat{H}_0$. In addition to the Bose-Hubbard terms, we consider the contribution from the NNN hoppings
\begin{equation}
\hat H_{\rm NNN}(t)
\! \! 
  =  \! \! 
\sum_{n=1}^{N_\rho}
\left[
i\rho_n(t)\,\hat a_{n+1}^\dagger \hat a_n
-
i\rho_n^*(t)\,\hat a_n^\dagger \hat a_{n+1}
\right],
\label{eq:NNNsecondquantized}
\end{equation}
so that the full many-body Hamiltonian is $\hat H_{\rm MB}(t)=\hat H_{\rm SSH}(t)+\hat H_{\rm NNN}(t)+\hat H_{\rm int}$, where $\hat H_{\rm SSH}(t) \equiv \hat H_{\rm 0,SSH}(t)$ for the simple SSH geometry and $\hat H_{\rm SSH}(t) \equiv \hat H_{\rm 1,SSH}(t)$ for the topological interface SSH geometry.

At the mean-field level, we restrict the many-body Bose-Hubbard dynamics to a
product-state ansatz~\cite{Trimborn08},
\begin{equation}
\label{eq:MFansatz}
|\Psi_{\rm MF}(t)\rangle
=
\frac{1}{\sqrt{M!}}
\left[
\sum_{n=1}^{N_A} \alpha_n(t)\hat a_n^\dagger
+
\sum_{n=1}^{N_B} \beta_n(t)\hat b_n^\dagger
\right]^M \! \! \! \! \!
|0\rangle .
\end{equation}
Here, $M$ is the number of particles in the SSH chain,
and the unit normalization $\sum_n|\alpha_n(t)|^2+\sum_n|\beta_n(t)|^2=1$ yields the effective mean-field nonlinearities  $g_{A,B}=U_{A,B}(M-1)$.

We now establish the central result of this paper: the NNN-assisted shortcut can
be extended exactly to the nonlinear mean-field regime by adding an independent
quadrature to the NNN couplings.  The resulting protocol preserves the
$A$-sublattice dark-state profile during the nonlinear evolution, despite the
presence of interactions. Remarkably, the two quadratures play distinct and
decoupled roles: the original NNN quadrature remains the same as in the linear
single-particle shortcut and cancels nonadiabatic transitions, whereas the
additional orthogonal quadrature compensates the mean-field interaction-induced phase
shifts. Importantly, this nonlinear shortcut is not perturbative in the interaction strength: within the mean-field description, the protocol exactly compensates the interaction-induced phase shifts and therefore remains valid, in principle, even for strong nonlinearities, provided the mean-field approximation applies. For the regimes considered below, the additional
quadrature remains well behaved and exhibits favorable scaling properties,
supporting the feasibility of an experimental implementation with
phase-tunable NNN hoppings. In the linear interaction-free context, phase-tunable hoppings have also been
recognized as a useful resource for optimizing topological excitation
transmission in dimerized lattices with long-range hopping~\cite{Zheng2024PhaseFactor}.
In the nonlinear setting considered here, the additional hopping quadrature
instead compensates the nonlinear phase shifts generated by interactions.


\subsection{Dark-subspace condition in the mean-field regime}

With the mean-field ansatz~\eqref{eq:MFansatz}, the interacting Bose-Hubbard dynamics is described by
the nonlinear Schrödinger equation
\begin{eqnarray}
i \dot \alpha_n \! \! \!
&=& \! \! \!
t_2 \beta_n \! + \! t_1 \beta_{n-1} \!
- \!i\rho_n^* \alpha_{n+1}
\! + \! i \rho_{n-1} \alpha_{n-1} \!\nonumber\\ 
\! &+& \!  g_A |\alpha_n|^2 \alpha_n, 
\label{eq:mf_an}
\\
i \dot \beta_n \! \! \!
&=& \! \!
t_2 \alpha_n + t_1 \alpha_{n+1}
+ g_B |\beta_n|^2 \beta_n .
\label{eq:mf_bn}
\end{eqnarray}

A useful starting point is to identify which parts of the linear dark-state construction survive in the nonlinear dynamics. Indeed, by assuming that the $B$ sublattice is kept empty, i.e. $\beta_n=0$ for $1 \leq n \leq N_B$, Eq.~\eqref{eq:mf_bn} reduces to the linear  condition~\eqref{eq:simple_dark_state_condition} determining the edge-mode structure. Despite the presence of mean-field interactions, the instantaneous dark-state profile is therefore unaltered. This result holds for both the simple and the topological-interface SSH chains. Interaction effects appear instead in the phase evolution required to keep this profile on the \(A\) sublattice. In the following, we denote \(g:=g_A\).

We now show that the dark-subspace constraint operates as a gauge-fixing condition. At the level of population transfer alone, one might be tempted to exploit a
phase freedom and replace the instantaneous real dark-state profile $a_n(t) \in \mathbb{R}$
by the locally phased profile
\begin{equation}
  \alpha_n(t)=e^{-i\phi_n(t)} a_n(t).
  \label{eq:local_phase_trial}
\end{equation}
Indeed, this transformation leaves the populations unchanged, and therefore does not affect the instantaneous density profile. From this
point of view, the phases \(\phi_n(t)\) represent a gauge freedom for protocols
whose objective is only to reproduce population transfer.

However, this freedom is lost if one requires the trajectory to remain exactly
inside the dark subspace. In that case the \(B\)-sublattice amplitudes must
satisfy at all times $\beta_n(t)=0$.
The equation of motion on the \(B_n\) site then imposes
$t_2 \alpha_n + t_1 \alpha_{n+1}=0$ which, together with the dark-state condition~\eqref{eq:simple_dark_state_condition} for $a_n$, immediately yields $\phi_{n+1}(t)=\phi_n(t)  \pmod{2\pi}$. Therefore the gauge freedom is restricted to a global phase, namely 
\begin{equation}
\label{eq:ansatz}
\alpha_n(t)=e^{-i\chi(t)} a_n(t)
\end{equation} 
Thus, one cannot compensate nonlinear interactions by a simple local dephasing while keeping the trajectory on the prescribed instantaneous dark-state mode. 

\subsection{Mean-field shortcut equation and necessity of an additional NNN quadrature}

We therefore seek NNN couplings $\rho_n(t)$ that compensate the mean-field-induced phase shifts while keeping the state on the dark-state trajectory. Precisely, we determine $\rho_n(t)$ such that the ansatz~\eqref{eq:ansatz} becomes an exact solution of the nonlinear Schr\"odinger equation. By inserting this ansatz into the relevant \(A\)-sublattice equation~\eqref{eq:mf_an}, one obtains
\begin{equation}
  i\dot a_n
  =
  -i\rho_n^*a_{n+1}
  +i\rho_{n-1}a_{n-1}
  +
  \left(g a_n^2-\dot\chi\right)a_n .
  \label{eq:central_mf_shortcut}
\end{equation}
This is the central mean-field inverse-engineering equation. Compared with the
linear shortcut equation, the only new contribution is the real nonlinear term
\((g \, a_n^2-\dot\chi)a_n\), which represents the local self-phase modulation
after accounting for the global phase. The global phase rate $\dot{\chi}(t)$ is determined self-consistently, and the \(B\) sublattice remains empty by construction.

A simple argument shows immediately why real-valued NNN couplings cannot compensate mean-field interaction effects during the edge state transport. Assuming $\rho_n(t) \in \mathbb{R}$, as the edge mode $a_n(t)$ is real-valued, the NNN contribution $-i\rho_n a_{n+1}
  +i\rho_{n-1}a_{n-1}$ in Eq.~\eqref{eq:central_mf_shortcut} corresponds to a purely imaginary term. In contrast, the mean-field term is real-valued. Thus, with real-valued NNN couplings, the nonlinear Schr\"odinger equation could only be satisfied with a spatially uniform dark state density $|a_n(t)|=a_0$ with an appropriate choice for $\chi(t)$. This corresponds to the very specific degenerate case $\epsilon=|t_2/t_1|=1$. Such a profile is not sufficient to transport the edge mode from the left end
to the right end of the chain, since transfer requires sweeping the localization
from \(|\epsilon|\ll1\) to \(|\epsilon|\gg1\). Thus real-valued NNN couplings
cannot, in general, provide an exact mean-field shortcut.

\subsection{Exact nonlinear shortcut with two control quadratures}

The limitation found above can be overcome by allowing the NNN-assisted
couplings to acquire a second quadrature. We decompose the NNN couplings into real and imaginary parts as
\begin{equation}
\label{eq:complexNNNcouplings}
  \rho_n(t)=\rho_n^R(t)+i\rho_n^I(t),
  \qquad
  \rho_n^R(t),\rho_n^I(t)\in\mathbb R .
\end{equation}
Upon restricting the dynamics to the mean-field ansatz~\eqref{eq:MFansatz}, the second-quantized
NNN Hamiltonian~\eqref{eq:NNNsecondquantized} contributes to the equations of motion for the
amplitudes $\alpha_n$ and $\beta_n$ through the single-particle operator $\hat H_{\rm NNN}^{(1)}(t)$
already introduced in Eq.~\eqref{eq:CDHamiltonian}. This NNN contribution remains linear and Hermitian
for arbitrary complex-valued $\rho_n(t)$. The introduction of a second quadrature corresponds to allowing the Hermitian NNN hopping to acquire a controllable complex phase. The real quadrature
corresponds to the interaction-free NNN-assisted shortcut
of Sec.~\ref{sec:linearshortcut}, while the additional imaginary quadrature
enables compensation of mean-field interaction-induced
phase shifts.

Phase-tunable hopping terms are within experimental reach, as illustrated by recent cold-atom implementations
of SSH-type tight-binding Hamiltonians~\cite{Gadway19,Ronco26}. More specifically, complex NNN hopping amplitudes,
including purely imaginary ones, have been realized
experimentally in momentum-space zigzag lattices,
with independent control over individual hopping
amplitudes and phases~\cite{An2018}.
A complementary approach exploiting quantum resonances and optimal control enables independent tuning of the real and imaginary parts of NN hopping amplitudes and has experimentally demonstrated NNN hopping in Floquet-engineered momentum-space lattices~\cite{Ronco26}.

One then substitutes the complex-valued NNN coupling~\eqref{eq:complexNNNcouplings} into the central mean-field shortcut equation~\eqref{eq:central_mf_shortcut}. By taking the imaginary part, using the fact that \(a_n(t)\) is real-valued, one finds that $\rho_n^R(t)$
is still given by the interaction-free NNN-coupling equation~\eqref{eq:linear_inverse_engineering_simple}. Thus the real quadrature \(\rho_n^R\) is determined by the usual
reverse-engineering procedure and cancels the nonadiabatic transitions that
would otherwise drive the state away from the instantaneous dark-state profile. On the other hand, the real part of this equation
yields the imaginary NNN quadrature $\rho_n^I(t),$ which compensates mean-field interactions
\begin{equation}
  \rho_n^I a_{n+1}
  +
  \rho_{n-1}^I a_{n-1}
  =
  \left(g \, a_n^2-\dot\chi\right)a_n .
  \label{eq:rhoI_nonlinear_recurrence}
\end{equation}
Here, $1\leq n\leq N_A$, with the boundary conditions
$\rho_0^I=\rho_{N_A}^I=0$. This second equation determines the additional quadrature \(\rho_n^I\), together
with the global phase rate \(\dot\chi\). Its role is to compensate the
site-dependent nonlinear self-phase modulation induced by the mean-field
interaction. In practice, $\rho_n^I(t)$ and $\dot\chi(t)$ are
determined simultaneously by solving the corresponding
linear system at each time.

Equations~\eqref{eq:linear_inverse_engineering_simple}
and
\eqref{eq:rhoI_nonlinear_recurrence} show that the two inverse-engineering
problems, respectively associated with the cancellation of nonadiabatic transitions and compensation of mean-field interactions, decouple. The real quadrature \(\rho_n^R\) implements the same linear
shortcut as before, suppressing nonadiabatic transitions to the bands. The
imaginary quadrature \(\rho_n^I\) is instead devoted to cancelling the nonlinear
mean-field phase shifts. Thus the exact nonlinear shortcut is obtained by
adding one independent control quadrature to account for self-phase modulation, without modifying the linear
reverse-engineering structure.

\subsection{Nonlinear shortcut in a simple and in a topological SSH chain}

\begin{figure}[htbp]
    \centering
    \includegraphics[width=8cm]{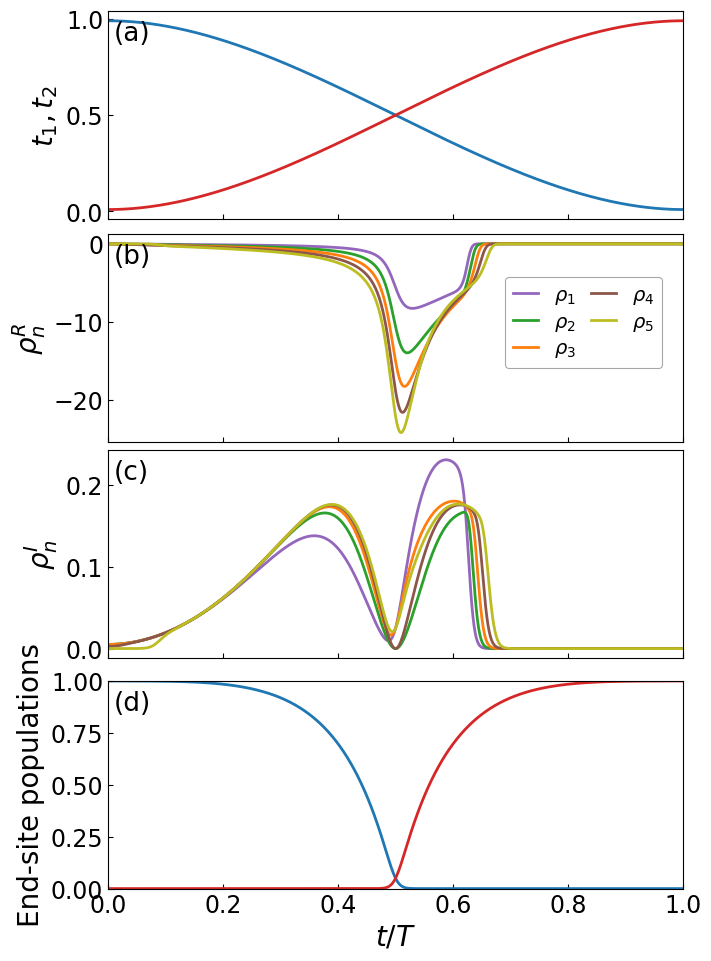}
    \caption{
    Nonlinear NNN-assisted shortcut for state transfer in a simple SSH chain with $N=39$ sites.
(a) Prescribed nearest-neighbor couplings $t_1(t)$~(blue) and $t_2(t)$~(red) as a function of time, defining the adiabatic passage between the two end sites. 
(b) Real quadratures $\rho_n^R(t)$ of the NNN couplings, which coincide with the linear inverse-engineered shortcut.
(c) Imaginary quadratures $\rho_n^I(t)$ for the considered mean-field regime.
Panels (b,c) show the first five NNN couplings, using the same color code indicated in panel (b). (d) Populations on the left~(blue) and right~(red) end sites as a function of time, showing near-perfect transfer.
The NN couplings are prescribed as $t_1(t)=t_0 [1-P_\delta(t/T)]$ and $t_2(t)=t_0 P_\delta(t/T)$, with
$P_\delta(x)=\delta+(1-2\delta)(3x^2-2x^3)$.
Parameters: $T=10/ t_0$, $g=0.5 \, t_0$, and $\delta=0.01$.
    }
    \label{fig:mf_simple}
\end{figure}

\begin{figure}[htbp]
    \centering
    \includegraphics[width=8cm]{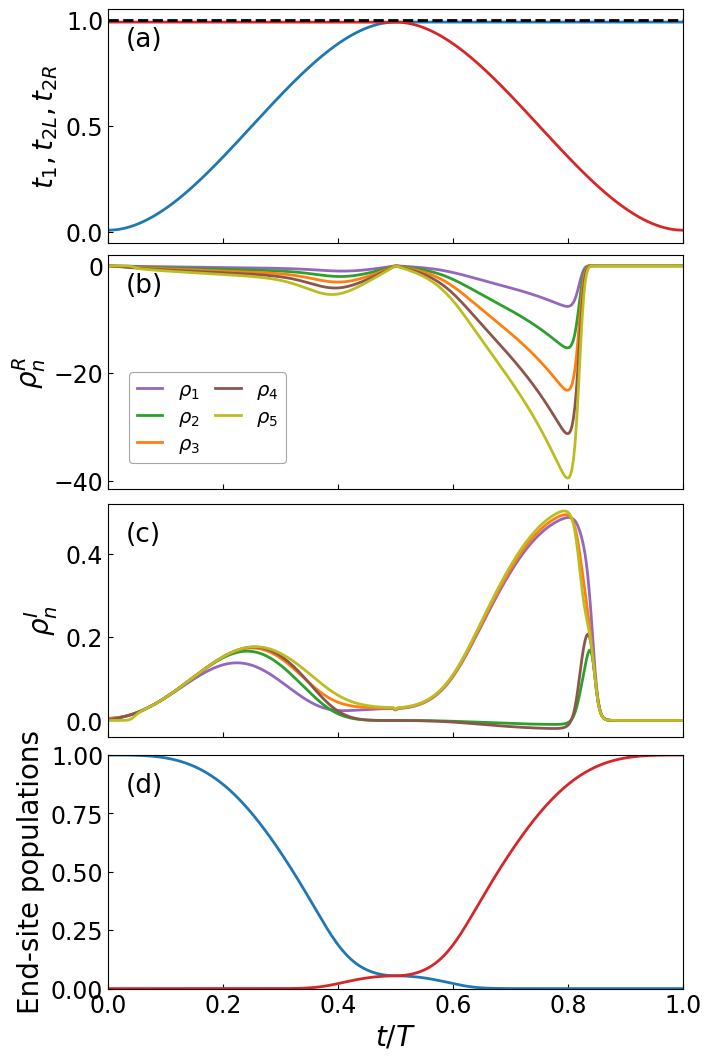}
    \caption{
    Nonlinear NNN-assisted shortcut for state transfer through a topological SSH interface.
(a) Prescribed NN couplings $t_1(t)$~(dashed black), $t_{2L}(t)$~(blue), and $t_{2R}(t)$~(red), defining the adiabatic passage across the interface. 
(b) Real quadratures $\rho_n^R(t)$ of the NNN couplings, which coincide with the linear inverse-engineered shortcut.
(c) Imaginary quadratures $\rho_n^I(t)$ for the considered mean-field regime. Panels (b,c) show the first five NNN couplings, using the same color code indicated in panel (b).
(d) Populations on the left~(blue) and right~(red) end sites as a function of time, showing near-perfect transfer across the interface.  The NN couplings are prescribed as $t_1(t)=t_0$, $t_{2L}(t)=t_0P_\delta(2t/T)$, and $t_{2R}(t)=t_0P_\delta(1)$ for $0\leq t\leq T/2$, and extended to the full interval by the mirror symmetry $t_{2L}(t)=t_{2R}(T-t)$ valid for $0 \leq t \leq T$.
The chain length, the function $P_\delta(x)$, and the remaining parameters are identical to Fig.~\ref{fig:mf_simple}.
    }
    \label{fig:mf_interface}
\end{figure}

To illustrate the nonlinear shortcut construction
and its connection with the interaction-free shortcut
reviewed in Sec.~\ref{sec:linearshortcut}, we
apply it to the two SSH geometries of
Fig.~\ref{fig:SSHscheme}. The results are shown
in Figs.~\ref{fig:mf_simple} and
\ref{fig:mf_interface}, respectively. In both cases we use chains with
\(N=39\) sites, total duration $T=10/t_0$, and mean-field interaction strength
$g=0.5 \, t_0$. For these parameters, the simultaneous presence of both NNN quadratures $\rho_n ^R(t)$ and $\rho_n^I(t)$ is necessary to ensure a high-fidelity quantum-state transfer. Panels (a) show the prescribed NN couplings defining the underlying adiabatic
passage. Panels (b) show the real NNN quadratures
$\rho_n^R(t)$, which coincide with the
interaction-free shortcut couplings derived
in Sec.~\ref{sec:linearshortcut} from
Eq.~\eqref{eq:linear_inverse_engineering_simple}.
These couplings suppress nonadiabatic transitions,
while the additional imaginary quadratures
$\rho_n^I(t)$, shown in panels (c), compensate
the nonlinear mean-field self-phase modulation. Finally, panels (d) show the resulting end-site
populations, demonstrating near-perfect nonlinear state transfer. In particular, the time-dependent end-site populations obtained in the
interacting mean-field regime with the full complex-valued NNN shortcut
coincide with those of the interaction-free dark-state trajectory. In this
sense, the imaginary NNN component restores the original dark-state transport
trajectory, up to a global phase. This property holds for both considered SSH geometries and is independent of the specific dark-state profiles.


For the midpoint-symmetric NN drives used here, namely $t_1(t)=t_2(T-t)$ for the simple SSH chain and \(t_{2L}(t)=t_{2R}(T-t)\) for the topological-interface SSH chain, the complex-valued NNN couplings inherit the
corresponding mirror symmetry. Explicitly, for both geometries, $\rho^{R,I}_{N_\rho+1-n}(t)=\rho^{R,I}_n(T-t)$, with $N_\rho$ the number of NNN couplings defined above. The role of the imaginary component is strongest when the density profile is
spatially nonuniform. Conversely, around \(t=T/2\), the dark state becomes
delocalized and the \(A\)-sublattice populations are nearly uniform
(\(|\epsilon|\simeq 1\)). In this regime, the nonlinear term \(g a_n^2\) is
approximately site independent and can therefore be absorbed into the global
phase rate \(\dot\chi\). This explains the dip of the imaginary NNN couplings
near the midpoint of the protocol.

\begin{figure}[htbp]
    \centering
    \includegraphics[width=8cm]{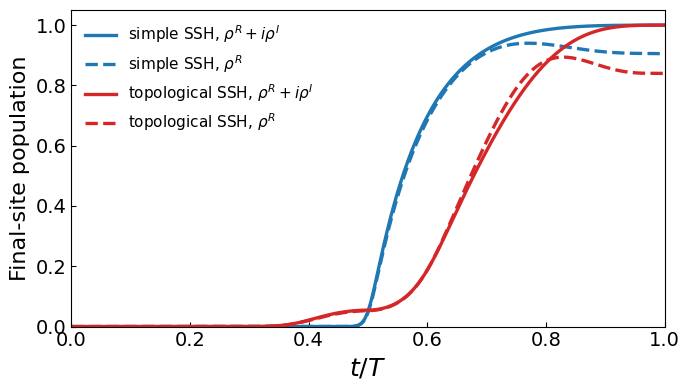}
    \caption{ Full nonlinear vs real-valued shortcut in
the nonlinear mean-field dynamics: right-end population of the simple~(blue) and topological-interface~(red) SSH chain as a function of the rescaled time. Solid lines correspond to the full complex
NNN couplings \(\rho_n(t)=\rho_n^R(t)+i\rho_n^I(t)\), while dashed lines
correspond to the real-valued shortcut \(\rho_n(t)=\rho_n^R(t)\). In both geometries, adding the imaginary component restores
near-perfect final-site population. Parameters identical to Figs.~\ref{fig:mf_simple},\ref{fig:mf_interface}.
    }
    \label{fig:mf_comparison}
\end{figure}

To isolate the role of the imaginary NNN component, Fig.~\ref{fig:mf_comparison}
compares the final-site population obtained with the full complex shortcut
\(\rho_n=\rho_n^R+i\rho_n^I\) and with the real-valued shortcut
\(\rho_n=\rho_n^R\), both propagated under the nonlinear mean-field dynamics. While the full complex shortcut yields a near-perfect quantum transfer ($1-\mathcal F \leq 10^{-7}$), the real-valued NNN couplings alone result in degraded fidelities, namely  \(\mathcal F \simeq 0.91\) and \(\mathcal F \simeq 0.84 \) for the simple and topological-interface SSH chains respectively. This
confirms that the real-valued NNN couplings implement the fast transport of the
dark-state profile, while the imaginary component is essential to compensate
the interaction-induced phase mismatch.

The two quadratures have distinct scaling properties with the protocol duration $T$. This can be seen directly from Eqs.~(\ref{eq:linear_inverse_engineering_simple},\ref{eq:rhoI_nonlinear_recurrence}) by writing the prescribed dark-state trajectory~\eqref{eq:ansatz}, or equivalently the prescribed NN amplitudes $t_{1,2}(t)$, as functions of the reduced time $s=t/T$. The NN amplitudes remain of order $t_0$ for any value of $T$, since they determine the instantaneous dark-state profile through the ratio $t_2/t_1$. The real NNN quadrature $\rho_n^R$, however, cancels the time derivative of this profile and thus scales as the inverse of the transfer time. By inspection of Eq.~\eqref{eq:linear_inverse_engineering_simple}, $\rho_n^R\sim 1/T$, so that this quadrature vanishes in the adiabatic limit~\cite{Dangelis20}. By contrast, the imaginary quadrature $\rho_n^I$ compensates the interaction-induced mean-field phase shifts, of order $g \: |a_n|^2$. Its leading contribution is therefore controlled by the nonlinear scale $g$ and by the instantaneous density profile, rather than by $1/T$. Under a time rescaling, $\rho_n^I(t)$ is shifted along the accelerated trajectory but is not multiplied by the speed factor, so that the relative overhead associated with interaction compensation decreases as the protocol is accelerated. Without this component, the uncompensated nonlinear phase error accumulates over the protocol duration and scales as $g| a_n|^2T$. Thus, $\rho_n^R(t)$ is the speed-dependent counterdiabatic resource, whereas $\rho_n^I(t)$ is the interaction-compensating resource. 

Interestingly, the additional interaction-compensating quadrature
does not modify the geometric speed of the prescribed mean-field
trajectory. Indeed, the instantaneous Fubini--Study speed,
defined as $v_{\rm FS}(t)=
\sqrt{\langle\dot\psi|\dot\psi\rangle
-|\langle\psi|\dot\psi\rangle|^2}$,
reduces here to $v_{\rm FS}(t)=\sqrt{\sum_n\dot a_n^2(t)}$
for the exact solution $|\psi(t)\rangle=e^{-i\chi(t)}\sum_n a_n(t)|A_n\rangle$.
The instantaneous Fubini--Study
speed is thus independent of the interaction strength $g$.
Since the Fubini--Study speed equals the instantaneous
energy uncertainty of the effective mean-field Hamiltonian~\cite{Anandan1990},
this uncertainty is also unchanged by interactions.
Thus, interactions modify the control resources required
to realize the prescribed trajectory, but neither its
geometric length nor its degree of saturation of the
Mandelstam--Tamm bound~\cite{Mandelstam1945}. Determining the minimum achievable transfer time under bounded
control resources, however, requires a separate
constrained optimization, as illustrated by recent
quantum-brachistochrone studies in noninteracting
qubit arrays~\cite{Stepanenko2025}.

\subsection{Mean-field shortcut in the many-body regime}

\begin{table}
\centering
\caption{
Full \(M\)-particle Bose-Hubbard propagation with the real-valued linear
shortcut and the full complex mean-field shortcut in simple SSH chains of
length \(N\), for fixed effective mean-field interaction strength
\(g=U_A(M-1)=0.5 \, t_0\).
}
\label{table:manybodyshortcut}
\begin{tabular}{cccccc}
\hline\hline
\(N\) & \(M\) & \(D\) & \(U_A/t_0\) & \(\mathcal F_{\{\rho_n^R\}}\) &
\(\mathcal F_{\{\rho_n^R+i\rho_n^I\}}\) \\
\hline
9  & 2 & 45   & 0.5000 & 0.5569 & 0.7508 \\
9  & 3 & 165  & 0.2500 & 0.5275 & 0.8164 \\
9  & 4 & 495  & 0.1667 & 0.4703 & 0.8386 \\
9  & 5 & 1287 & 0.1250 & 0.4132 & 0.8495 \\
9  & 6 & 3003 & 0.1000 & 0.3611 & 0.8560 \\
\hline
19 & 2 & 190  & 0.5000 & 0.6085 & 0.7811 \\
19 & 3 & 1330 & 0.2500 & 0.5819 & 0.8376 \\
19 & 4 & 7315 & 0.1667 & 0.5291 & 0.8568 \\
\hline
39 & 2 & 780   & 0.5000 & 0.6260 & 0.7929 \\
39 & 3 & 10660 & 0.2500 & 0.5990 & 0.8464 \\
\hline\hline
\end{tabular}
\end{table}

We now investigate the performance of the mean-field shortcut in the genuine many-body Bose-Hubbard dynamics for an initial state $| \Psi(0) \rangle = \tfrac{1}{\sqrt{M!}} a_1^{\dagger M}|0 \rangle$ localized at the left end of a simple SSH chain. 
We quantify the transfer performance by the final-state
fidelity
$\mathcal{F}=|\langle\Psi_{\rm tar}|\Psi(T)\rangle|^2$,
where
$|\Psi_{\rm tar}\rangle=
(\hat a_{N_A}^\dagger)^M|0\rangle/\sqrt{M!}$
corresponds to all $M$ bosons occupying the
right-end site. \enlargethispage{\baselineskip} For the simple SSH chain considered
here, $N_A=N_0$. The results are summarized in Table~\ref{table:manybodyshortcut}. Keeping the effective mean-field interaction strength \(g=U_A(M-1)=0.5\) fixed, we propagated \(M=2,\ldots,6\) bosons in the simple
SSH chain. {The dimension~$D$ of the $M$-particle Hilbert space increases quickly with the particle number
and limits our benchmark to moderate values of $M$. The full complex mean-field shortcut systematically outperforms the
real-valued shortcut, showing that the imaginary NNN component remains useful
beyond the mean-field approximation. Moreover, the efficiency of the full
nonlinear shortcut improves as the particle number increases, although it
appears to saturate below unity for the considered particle numbers. In
contrast, the real-valued linear shortcut worsens as
\(M\) increases.} 

This residual discrepancy reflects the fact that the exact many-body dynamics
explores regions of Hilbert space beyond the single-wave-function product-state manifold. Interactions generate correlations and depletion that are absent from
the mean-field construction.

\section{Pontryagin-based optimization in the full many-body regime}
\label{sec:PMPmanybody}

We now use Pontryagin-based optimization~\cite{PontryaginBook,Sugny21,DavidPRX21,Ansel2024QOC} to address quantum-state transfer beyond the mean-field regime. 

As a preliminary benchmark, we first show that restricted-control PMP protocols provide efficient transfer schemes in the linear, noninteracting regime, requiring fewer independently controlled NNN couplings than the exact shortcut construction. The main objective of this section, however, is to extend high-fidelity transfer to the genuine many-body Bose-Hubbard dynamics. 

An optimization over real-valued NNN couplings substantially improves the transfer but saturates below unity.
Guided by the mean-field shortcut construction, which identified the imaginary NNN component as the quadrature compensating interaction-induced phase shifts, we address the full many-body problem with an enlarged control manifold enabling phase-tunable NNN couplings. While we use  the same physical control resource as in the mean-field shortcut problem, namely individually addressed phase-tunable NNN hoppings, we switch from trajectory engineering in a low-dimensional mean-field manifold to optimal control in the full many-body Hilbert space. By relaxing the requirement to follow the dark-state zero mode, we obtain a near-unity transfer fidelity in this regime. This demonstrates that the imaginary NNN quadrature remains an important control resource for compensating interaction effects beyond the mean-field description.

\subsection{PMP optimization strategy}

The role and motivation for PMP optimization are qualitatively different in the noninteracting and interacting regimes. 

In the noninteracting case, exact shortcut protocols already provide perfect transfer by enforcing the evolution along the instantaneous dark-state trajectory. The purpose of PMP is therefore mainly practical: it relaxes the requirement of independently controlling all NNN couplings and searches for high-fidelity transfer within a reduced control manifold. The optimized trajectory is no longer constrained to follow the adiabatic edge mode, but this additional freedom results in a simpler and more experimentally accessible control structure.

In the interacting many-body regime, the situation is more fundamental. The mean-field shortcut provides an exact solution only within the mean-field ansatz, and its direct application to the full Bose-Hubbard dynamics does not yield sufficiently reliable quantum transfer. PMP optimization is then used not merely to simplify an already exact protocol, but to find a genuinely many-body trajectory in the full Fock space. In this regime, relaxing the dark-state-following constraint becomes essential for recovering high transfer fidelity.

The optimization minimizes the final transfer
infidelity, namely $J[\{\rho_n(t)\}]=1-\mathcal F$ with $\mathcal F=\left|\langle \Psi_{\rm tar}|\Psi(T)\rangle\right|^2$,
over a set of NNN control functions. In the full many-body problem, $|\Psi(t)\rangle$ denotes the Bose-Hubbard state propagated in the fixed-$M$ Fock sector, while in the non-interacting regime the fidelity is obtained by taking the single-particle wave-function $|\psi(t)\rangle$. Technical details of the optimization procedure can be found in Appendix~A. 

 In all PMP optimizations considered below, the nearest-neighbor SSH couplings $t_1(t)$ and $t_2(t)$ for the simple SSH chain, or $t_1(t)$, $t_{2L}(t)$, and $t_{2R}(t)$ for the
topological-interface SSH chain, are kept fixed to the prescribed smooth ramps used in the shortcut protocols. These correspond to the NN hoppings of Fig.~\ref{fig:mf_simple} for the simple SSH chain and of Fig.~\ref{fig:mf_interface} for the topological-interface SSH chain. The PMP optimization thus affects only the NNN
hoppings $\{\rho_n(t)\}_{n=1}^{N_\rho}$, subject to the specific parametrizations and
constraints described below.


\subsection{Restricted-control PMP in the noninteracting regime}
\label{subsec:noninteractingPMP}

We first use Pontryagin-based optimization~\cite{PontryaginBook,Sugny21,DavidPRX21,Ansel2024QOC} to construct simplified transfer protocols in the noninteracting regime. Although exact, the shortcut protocol requires an independent control of several NNN couplings. Such individual addressing may become experimentally demanding, whereas grouped controls, in which several couplings share the same time dependence, may be more practical -- for instance enabling the implementation of multiple couplings with a single laser field. The broader relevance of reduced-control strategies for QST is also illustrated by recent work on quasicrystal Fibonacci-chain pumps, where state transfer between distant qubits is achieved through minimal endpoint control~\cite{Ghosh2025}. The results, summarized here and detailed in Appendix~A, are qualitatively different for the two geometries of the SSH chain.

For the simple SSH chain, efficient QST is obtained with a single global NNN control, $\rho_n(t)\equiv \rho(t),
\qquad 1\leq n\leq N_0-1$. In contrast with the shortcut trajectory, the propagated state can acquire a significant population on the (B) sublattice. The resulting protocol yields near-perfect single-particle transfer $\mathcal F \gtrsim 0.9995$ in the simple SSH chain and remains robust against moderate control imperfections and static off-diagonal disorder.

For the topological-interface SSH chain, our optimization procedure indicates that a single global NNN control is no longer sufficient. We instead group the NNN couplings into three controls, $\rho_L(t)$, $\rho_M(t)$, and $\rho_R(t)$, associated respectively with the left segment, the interface coupling, and the right segment:
\begin{equation}
\rho_n(t)=
\begin{cases}
\rho_L(t), & 1\leq n\leq N_0-1,\\[1mm]
\rho_M(t), & n=N_0,\\[1mm]
\rho_R(t), & N_0+1\leq n\leq 2N_0-1.
\end{cases}
\label{eq:grouped_controls}
\end{equation}
Using a finite set of Fourier components, we obtain near-unit transfer fidelities, $\mathcal F \gtrsim 0.999$, for an SSH chain with a topological interface and ($N=4N_0-1=39$) sites. Unlike in the simple-chain benchmark, the optimized NNN controls are large compared with the NN hoppings. They therefore open an effective same-sublattice transport channel and strongly suppress leakage to the (B) sublattice, even though the time-dependent state differs significantly from the moving edge mode used in the shortcut construction.

Thus, in the single-particle sector, PMP trades dark-state following for a
simpler implementation protocol relaxing the need for individual addressing. Compared with the simple SSH chain, the presence of a topological interface
therefore requires a richer reduced-control manifold and leads to larger,
more rapidly oscillating optimized controls. This large amplitude and the fast oscillations of the control depicted in Fig.~\ref{fig:TopologicalSSHChainControl} are the price to pay for seeking a near-perfect transfer with a drastically reduced control manifold.

\subsection{Many-body PMP with phase-tunable NNN hoppings}

We now return to the case where the NNN couplings can be addressed
individually, and optimize the transfer directly in the full many-body
Bose-Hubbard Hilbert space. Guided by the mean-field analysis, we allow each
NNN hopping to be complex-valued, i.e. $\rho_n(t)=\rho_n^R(t)+i\rho_n^I(t),
\qquad 1\leq n\leq N_\rho=2N_0-1$.  For the topological-interface chain with $N_0=5$, this corresponds to $N=19$ sites, and $2N_0-1=9$ NNN couplings and thus $18$ independent controls. 

To obtain a tractable control problem, we consider an SSH chain with a topological interface loaded with (M=2) bosons, of total length $N=4 N_0-1=19$ sites. The initial state is chosen with both bosons occupying the left edge site, $|\Psi(0)\rangle=
\frac{1}{\sqrt{2}}
\left(\hat a_1^\dagger\right)^2|0\rangle $, and the target state corresponds to a full boson transfer to the right edge site,
$
|\Psi_{\rm tar}\rangle=
\frac{1}{\sqrt{2}}
\left(\hat a_{2N_0}^\dagger\right)^2|0\rangle .
$
The PMP optimization is performed directly in the fixed-$M$ Bose-Hubbard Fock sector, without assuming a mean-field product-state ansatz as in the shortcut procedure.

\begin{figure}[htbp]
    \centering
    \includegraphics[width=8cm]{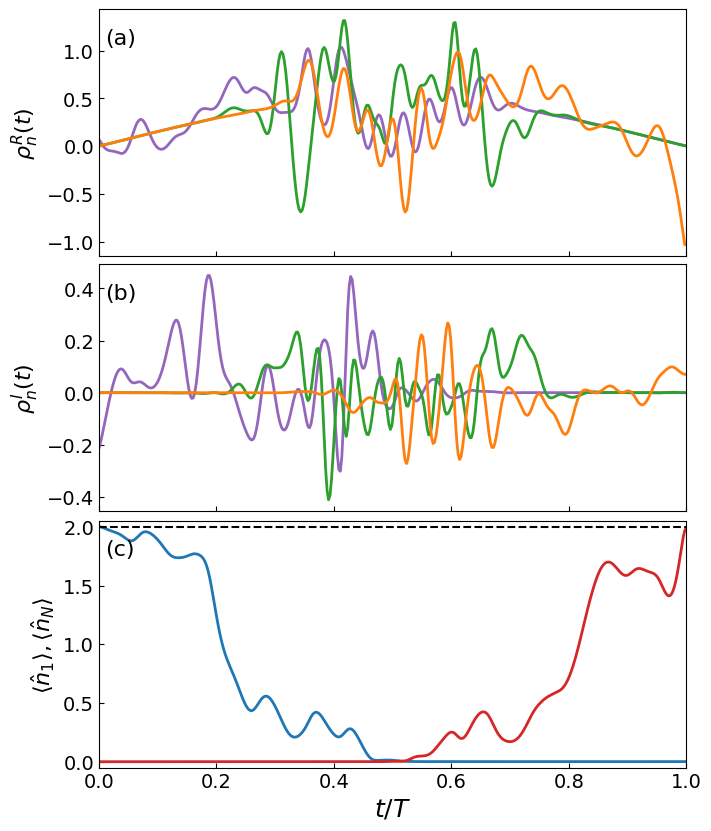}
    \caption{Many-body PMP protocol with complex site-dependent NNN controls in an SSH
    chain with a topological interface. The optimization is performed directly
    in the fixed-\(M\) Bose-Hubbard Fock sector for \(M=2\) bosons.
(a) Real quadratures \(\rho_n^R(t)\) of selected NNN couplings:
    \(\rho_1^R\)~(purple), \(\rho_5^R\)~(green), and \(\rho_9^R\)~(orange).
    (b) Corresponding imaginary quadratures \(\rho_1^I\), \(\rho_5^I\), and
    \(\rho_9^I\), with the same color convention.
    (c) Time-dependent occupations of the initial and target edge sites,
    \(\langle \hat n_1\rangle\)~(site $A_1$, blue) and \(\langle \hat n_N\rangle\)~(site $A_{2 N_0}$, red). The dashed
    horizontal line indicates the total particle number \(M=2\). Topological-interface SSH chain with $N=19$ sites~($N_0=5$) and interaction strength $g=0.5 \, t_0$. The time profile of the NN hoppings $\{t_1(t),t_{2L}(t),t_{2R}(t)\}$ is prescribed as in Fig.~\ref{fig:mf_interface}. Total transfer time $T=10/t_0$. Optimization using $N_{\rm iter} \simeq 2.7 \times 10^3$ iterations with a variable control-update step $0 \leq \eta \leq 2$~(see Eq.~\eqref{eq:controlstepdefinition}).}
   \label{fig:manybody_complex_pmp}
\end{figure}

The resulting two-quadrature PMP protocol is shown in Fig.~\ref{fig:manybody_complex_pmp}. Panels~(a) and~(b) display three selected NNN couplings, corresponding to the left edge-side coupling, the interface coupling, and the right edge-side coupling. Both the real~($\rho_n^R(t)$) and the imaginary~($\rho_n^I(t)$) quadratures  exhibit oscillatory patterns at intermediate time scales, while remaining finite and numerically well behaved throughout the evolution. The optimized protocol therefore does not rely on singular or excessively large control amplitudes.

The corresponding two-body dynamics is shown in Fig.~\ref{fig:manybody_complex_pmp}(c), where we plot the average occupation of the initial and target edge sites. The transfer is nearly complete, with an average target-site final population $\langle \hat{n}_{N} \rangle(T) \simeq 1.99$ and a quantum fidelity $\mathcal F\simeq 0.995.$ The time dependence is clearly nonmonotonic and differs markedly from the smooth population transport expected in an adiabatic protocol or in the exact shortcut construction based on dark-state following. This indicates that the PMP algorithm exploits a genuinely two-body trajectory in the full Fock space, rather than remaining confined to the local low-dimensional manifold spanned by the edge states of the topological chain. 

This result should be compared with the optimization restricted to real-valued NNN couplings, which saturates at $\mathcal F\simeq0.855$ even when all site-dependent NNN couplings are independently controlled. The improvement to $\mathcal F\simeq0.995$ demonstrates that the imaginary NNN quadrature, identified analytically in the mean-field shortcut construction as the component compensating interaction-induced phase shifts, remains a crucial control resource in the full interacting Bose-Hubbard dynamics.

Finally, these high-fidelity QST protocols are not restricted to the specific SSH chain parameters considered above. Additional PMP optimization runs, whose details are omitted for compactness, show that high-fidelity QST with $\mathcal F\geq 0.995$ is also recovered for a longer topological-interface SSH chain, $N=39$, and for a stronger onsite interaction, $g/t_0= 1.0$.
In these more demanding cases, the main difference is a slower convergence of the optimization routine when using comparable optimization settings. Similar results are obtained in simple SSH chains. Thus, our numerical benchmarks suggest that the same phenomenology  persists for longer SSH chains or in the presence of stronger onsite interactions.

\section{Conclusion}

 Compensating unwanted dynamical effects is crucial in many quantum-information settings. Examples include the
compensation of light-shift-induced phase errors in
Rydberg-atom quantum gates~\cite{Saffman2005}, the use of
atomic interactions to compensate coherent qubit
errors~\cite{Arimundo2022}, and the design of robust
shortcut protocols for quantum-state transfer and
transport~\cite{Ruschhaupt2012,Mortensen2018}.
Here, we address a related problem: the compensation of interaction effects during the quantum-state transfer in topological SSH chains.  Our approach builds on an interaction-free shortcut protocol~\cite{Dangelis20} to design an exact nonlinear shortcut. A similar strategy has been developed for dissipative
quantum systems, where protocols designed in the absence of dissipation can be adapted to non-Hermitian dynamics
through suitable compensating controls~\cite{Impens2019}. 

We have first extended NNN-assisted shortcut-to-adiabaticity protocols in SSH chains to the interacting mean-field regime. We have shown that mean-field interactions can be compensated exactly by adding an orthogonal quadrature to the NNN couplings, while the original NNN quadrature remains unchanged from the interaction-free shortcut. In the mean-field dynamics, the cancellation of nonadiabatic transitions and the compensation of interaction-induced phase shifts are therefore handled independently by the two NNN quadratures. The resulting nonlinear shortcut is still generated by a Hermitian Hamiltonian, so that the additional quadrature corresponds to a controllable complex phase of the NNN hopping rather than to a non-Hermitian ingredient. These analytical results show that interactions do not simply degrade transfer fidelity; rather, they increase the dimensionality of the control manifold required to maintain near-perfect quantum-state transfer.

We have then tested this picture in the genuine many-body Bose-Hubbard dynamics. In this regime, the mean-field shortcut still provides a significant improvement over the real-valued interaction-free shortcut, but its fidelity saturates below unity because the exact many-body evolution explores correlations and depletion beyond the mean-field product-state ansatz. To overcome this limitation, we used Pontryagin-based optimization directly in the fixed-particle-number Fock space, focusing on the two-boson case in finite SSH chains for which the procedure remains numerically tractable. By optimizing simultaneously over the two orthogonal quadratures of the NNN hoppings, we restore near-perfect transfer fidelity in this genuine two-boson setting.

Our results identify the additional orthogonal NNN quadrature as a key control resource for compensating interaction-induced phase shifts. Conversely, this suggests that suitably engineered phase-tunable NNN couplings could be used to emulate nonlinear phase dynamics in otherwise linear SSH chains. More generally, the present work indicates that high-fidelity transfer in interacting topological chains can be recovered by enlarging both the control manifold and the set of accessible quantum pathways: in the mean-field regime, through additional orthogonal hopping quadratures that compensate interaction-induced phase shifts, and beyond mean field, through many-body optimal-control trajectories that are not restricted to dark-state following.

\noindent \emph{Acknowledgments.}
The authors are grateful to Natanael C. Costa and Felipe A. Pinheiro for discussions. We acknowledge support from the Institut Universitaire de France, and from the ANR project QuCoBEC (ANR-22-CE47-0008-02) and from the CAPES-COFECUB (20232475706P) program. F.I. acknowledges support from the Brazilian agencies CNPq (305638/2023-8) and FAPERJ (210.570/2024). 

\emph{Author contributions and AI-use disclosure} F.I. and D.G.-O. jointly conceived the project. F.I. developed the theoretical framework, performed the numerical simulations, analyzed the results, and wrote the manuscript. D.G.-O. suggested the many-body extension, contributed to the theoretical interpretation, to the discussion of the results, and to the writing and revision of the manuscript. Generative AI tools were used to assist with language editing and manuscript polishing, and to produce the graphical illustration accompanying the article on Quantum's website. All scientific content, calculations, simulations, interpretations, and final text were checked and approved by the authors, who take full responsibility for the work.

\appendix

\section*{APPENDIX A: Pontryagin maximum principle and linear restricted-control SSH protocols}

\subsection*{A.1 Basics of Pontryagin maximum principle optimization}

We briefly recall the Pontryagin-based procedure used in the numerical optimizations~\cite{PontryaginBook,Sugny21,DavidPRX21,Ansel2024QOC}. Consider a finite-dimensional Schrödinger equation
\begin{equation}
\label{eq:statepropagation}
i\frac{d}{dt}|\psi(t)\rangle = H[\mathbf u(t)]|\psi(t)\rangle ,
\end{equation}
where $\mathbf u(t)=(u_1(t),u_2(t),\ldots)$ denotes the set of control functions. The dimension of the Hilbert space depends on the problem under consideration: it is the single-particle Hilbert-space dimension in the noninteracting calculations and the fixed-$M$ Fock-space dimension in the many-body calculations.

 For a final-time objective only, the PMP adjoint, or costate, can be
represented by an auxiliary nonphysical state \(|\tilde\psi(t)\rangle\).
In the variational formulation, this state plays the role of a
time-dependent Lagrange multiplier enforcing the Schrödinger dynamics,
and carries the sensitivity of the final fidelity backward in time. It is
defined by the terminal condition
\begin{equation}
\label{eq:adjoint_terminal}
|\tilde\psi(T)\rangle =
 \langle \psi_{\rm tar}|\psi(T)\rangle
|\psi_{\rm tar}\rangle ,
\end{equation}
and is propagated backward in time according to
\begin{equation}
\label{eq:adjoint_propagation}
i\frac{d}{dt}|\tilde\psi(t)\rangle =
H[\mathbf u(t)]|\tilde\psi(t)\rangle .
\end{equation}

With this convention, the Pontryagin Hamiltonian reads
\begin{equation}
\mathcal H_P(t)
=
\operatorname{Im}
\left[
\langle \tilde\psi(t)|
H[\mathbf u(t)]
|\psi(t)\rangle
\right].
\end{equation}
The PMP stationarity condition for an unconstrained control reads
\begin{equation}
\frac{\partial \mathcal H_P}{\partial u_\alpha}(t)=0 ,
\end{equation}
or, more generally, the optimal control maximizes
\(\mathcal H_P\) over the admissible control set.

For the iterative numerical implementation, we use the corresponding gradient
\begin{equation}
\label{eq:gradientobtention}
G_\alpha(t)
=
\frac{\partial \mathcal H_P}{\partial u_\alpha}(t)
=
\operatorname{Im}
\left[
\left\langle \tilde\psi(t)\left|
\frac{\partial H}{\partial u_\alpha}
\right|\psi(t)\right\rangle
\right].
\end{equation}
The control is then updated according to
\begin{equation}
\label{eq:controlstepdefinition}
u_\alpha^{(k+1)}(t)
=
u_\alpha^{(k)}(t)
+
\eta\,G_\alpha^{(k)}(t),
\end{equation}
where the value of \(\eta\) may be adjusted during the optimization routine
so as to increase the final fidelity after each accepted update. The forward--backward propagation and update of the controls are repeated until the final fidelity has converged within a given tolerance threshold, or until a prescribed maximum number of iterations is reached.

In its simplest implementation, the PMP routine uses a direct time discretization into a finite number of time bins, here $N_{\rm bins}=100$, associated with a time step $\Delta t=T/N_{\rm bins}$. Each control is then represented by a vector of dimension $N_{\rm bins}$, so that the PMP update explores a control space of dimension $N_{\rm controls}\times N_{\rm bins}$. As a final step, by performing a linear interpolation over the optimized controls, one can validate the Schr\"odinger propagation by using a much larger number $N_{\rm bins-final} \sim 10^3-10^4$ of time bins than the one used to perform optimization.

\subsection*{A.2 PMP-optimized protocol for a simple SSH chain}

As a first restricted-control benchmark, we consider the simple SSH chain with a single global NNN control $\rho_n(t)\equiv\rho(t)$. The PMP optimization yields a final fidelity $\mathcal F \gtrsim 0.9995 $, showing that near-perfect transfer can be recovered even after reducing the control manifold from $N_0-1$ independent shortcut couplings to a single time-dependent function. 

\begin{figure}[htbp] 
\begin{center} \includegraphics[width=8 cm]{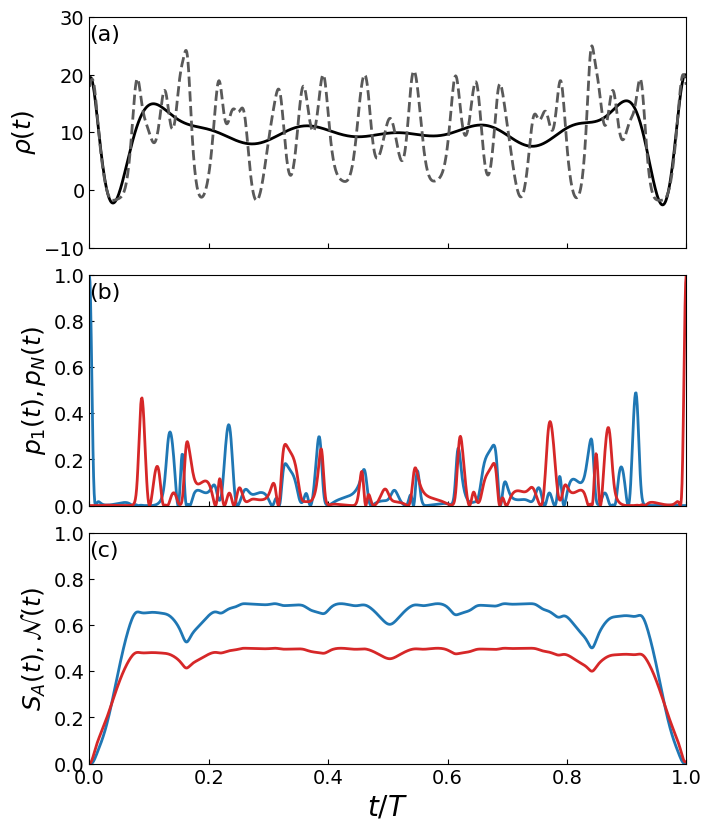} 
\end{center}
\caption{
Restricted-control PMP transfer in a simple linear SSH chain of $N=19$ sites~($N_0=10$) in the single-excitation regime.
(a) Optimized global NNN coupling $\rho_{\rm opt}(t)$~(gray dashed), with a smooth polynomial fit~(solid dark curve) used for propagation.
(b) Time-dependent populations at the left~(blue) and right~(red) ends of the chain.
(c) Sublattice entropy $S_A(t)$~(blue) and negativity $\mathcal{N}(t)$~(red) showing that the PMP trajectory transiently occupies both sublattices, unlike the shortcut trajectory which remains on the $A$ sublattice.
 The time profile of the NN hoppings $\{t_1(t),t_{2}(t)\}$ is prescribed as in Fig.~\ref{fig:mf_simple} with $T=10/t_0$. The optimization used $N_{\rm iter} \simeq 2 \times 10^4$ iterations with a variable control-update step $0 \leq \eta \leq 2$. The optimized protocol reaches $\mathcal F\simeq0.9998$ using a single global NNN control $\rho_n(t)\equiv\rho(t)$.}
\label{fig:SimpleSSHChainControl}
\end{figure}

 To characterize the PMP trajectory in Fig.~\ref{fig:SimpleSSHChainControl}(c), we use two sublattice indicators. Writing the single-particle state as
\[
|\psi(t)\rangle=\sum_{n=1}^{N_0}c_{2n-1}(t)|A_n\rangle+
\sum_{n=1}^{N_0-1}c_{2n}(t)|B_n\rangle ,
\]
we define the sublattice populations
\begin{eqnarray}
p_A(t) & = &\sum_{n=1}^{N_0}|c_{2n-1}(t)|^2, \nonumber \\
p_B(t) & = &\sum_{n=1}^{N_0-1}|c_{2n}(t)|^2=1-p_A(t). \nonumber
\end{eqnarray}
Equivalently, the state may be viewed as a bipartite state between the $A$ and $B$ sublattices, including the corresponding vacuum state on the unoccupied sublattice. The reduced density matrix then has eigenvalues $p_A(t)$ and $p_B(t)$, giving the sublattice entropy
\[
S_A(t)=-p_A(t)\log p_A(t)-p_B(t)\log p_B(t).
\]
The partial transpose has one negative eigenvalue $-\sqrt{p_A(t)p_B(t)}$; we therefore plot the associated negativity
\[
\mathcal{N}(t)=\sqrt{p_A(t)p_B(t)} .
\]
Both quantities vanish for a state supported on a single sublattice and become nonzero when the PMP trajectory transiently occupies both sublattices.

The resulting trajectory is qualitatively different from the inverse-engineered shortcut: the state does not remain locked to the instantaneous edge mode and transiently populates both sublattices, as shown in Fig.~\ref{fig:SimpleSSHChainControl}. This illustrates the basic mechanism by which optimal control trades trajectory simplicity for a reduced control manifold.

\subsection*{A.3 PMP-optimized protocol for an SSH chain with a topological interface}

\begin{figure} [htbp]
\begin{center}
\includegraphics[width=8cm]{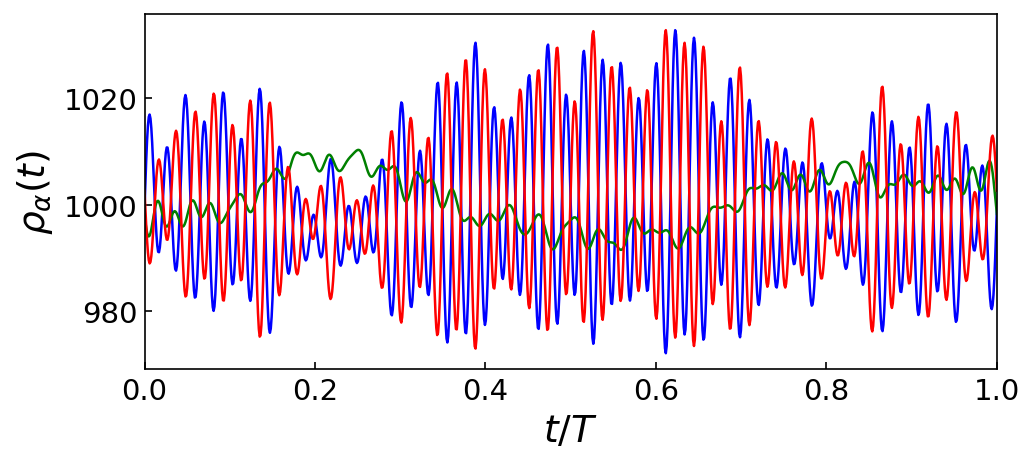}
\end{center}
\caption{Simplified NNN-assisted quantum-state transfer through a topological-interface SSH chain with $N=39$ sites~($N_0=10$),
optimized within a Fourier-restricted control manifold. The figure shows the
time-dependent profiles of the three grouped NNN couplings
\(\rho_L(t)\)~(blue), \(\rho_M(t)\)~(green), and \(\rho_R(t)\)~(red),
associated respectively with the left bulk, the interface coupling, and the
right bulk. The grouped controls are parametrized according to Eqs.~\eqref{eq:rho_alpha}--\eqref{eq:fixingcondition} with $N_{\rm modes}=50.$  The time profile of the NN hoppings $\{t_1(t),t_{2L}(t),t_{2R}(t)\}$ is prescribed as in Fig.~\ref{fig:mf_interface} with $T=10/t_0$. The optimization starts from a small random initialization
and reaches \(\mathcal{F}>0.999\) after $N_{\rm iter} \simeq 500$ iterations with a variable control-update step $0 \leq \eta \leq 2$.
}
\label{fig:TopologicalSSHChainControl}
\end{figure}

We now consider quantum-state transfer in the more involved case of an SSH
chain with a topological interface. To avoid addressing each NNN coupling
individually, we group them into three time-dependent amplitudes $\{ \rho_L(t),\rho_M(t),\rho_R(t) \}$, defined in Eq.\eqref{eq:grouped_controls}, corresponding
respectively to the left bulk, the NNN coupling at the interface, and the right bulk. The optimized controls $\rho_\alpha(t)$ with $\alpha=L,M,R$ are shown in Fig.~\ref{fig:TopologicalSSHChainControl}.


In this example, the PMP optimization based on a direct time-bin parametrization yields optimized controls with very sharp temporal variations. We therefore use instead a Fourier-restricted optimization, in which the controls are expanded over a truncated Fourier basis with harmonics up to $N_{\rm modes}$. The optimization is then performed directly on the Fourier coefficients, yielding smooth control functions by construction. In practice, the gradient with respect to each Fourier coefficient is obtained from the time-bin PMP gradient $G_\alpha(t)$ by the chain rule.


Precisely, we decompose the grouped left and right NNN controls into a constant scale and a modulated contribution as 
\begin{equation}
\label{eq:rho_alpha}
\rho_\alpha(t)=\rho_0[1+\delta_\alpha(t)],
\qquad \alpha=L,R,
\end{equation}
and optimize only over the two relative Fourier-restricted modulations $\delta_L(t)$ and $\delta_R(t)$ independently. Explicitly, the PMP routine optimizes over the set of coefficients $\{ \delta_{\alpha 0}, c_{\alpha n}, d_{\alpha n} \}$ entering their truncated Fourier expansion:
\begin{align}
& \delta_{\alpha}(t) \! \! = \! \! \delta_{\alpha 0} \! + \! \! \! \!  \!\sum_{n=1}^{N_{\rm modes}} \! \! \left[ c_{\alpha n} \cos \left(\frac {2 \pi n t} {T} \right) \! + \! d_{\alpha n} \sin \left(\frac {2 \pi n t} {T} \right) \right] \!  \nonumber \\
\end{align}
for $\alpha=L,R.$ By contrast, we choose to fix the interface control  $\rho_M(t)=\rho_0[1+\delta_M(t)]$ according to 
\begin{equation}
\label{eq:fixingcondition} \delta_M(t) = -\delta_L(t)-\delta_R(t) \,.
\end{equation}
A heuristic justification is that the interface control affects only a single dynamical coupling, in contrast to the grouped controls $\delta_{L},\delta_{R}$ affecting a full segment of the SSH chain. The validity of this restricted parametrization is confirmed a posteriori by the success of the numerical optimization, which explores a control space of dimension $N_{\rm controls} \times (2 N_{\rm modes}+1)$ with $N_{\rm controls}=2$.


The optimized protocol naturally identifies a different operating
regime: instead of transporting the excitation by adiabatically following the
conventional edge/interface modes, it uses strong next-nearest-neighbor
couplings to create an effective same-sublattice transport channel, yielding a
quantum fidelity \(\mathcal{F}\geq 0.999\).

In this regime, the occupation of the opposite sublattice remains strongly
suppressed, so that the nearest-neighbor leakage paths are dynamically
neutralized. The required controls are simple in structure: three bounded NNN
couplings oscillating around a common large value, as illustrated in
Fig.~\ref{fig:TopologicalSSHChainControl}. This simplicity comes with two
practical requirements. First, the common NNN coupling scale must be large
compared with the SSH couplings, \(\rho_{L,M,R} \sim 10^3 t_0 \). Second, the
relative modulations require a moderate Fourier bandwidth. In the example
shown, \(N_{\rm modes}=50\) is sufficient to reach \(\mathcal{F}>0.999\)
after $N_{\rm iter}=500$ iterations from a small random initialization.
When a smaller Fourier basis is used, however, the fidelity is degraded under the same optimization conditions: for instance, for \(N_{\rm modes}=40\) and \(N_{\rm modes}=30\), the optimized protocol delivers a much lower fidelity
\(\mathcal{F}\simeq 0.40\) and \(\mathcal{F}\simeq 0.20\). This rapid decrease suggests that 
intermediate-frequency Fourier components play an important role in the QST driving within this
restricted-control scheme.

 If $G_\alpha(t_j)$, $\alpha=L,M,R$, denotes the PMP gradient density associated with the grouped controls \(\rho_\alpha(t)\) on the numerical time grid $t_j$, the chain rule gives the Fourier-projected PMP gradients used to update the coefficients of the restricted parametrization. As a result of the fixing condition~\eqref{eq:fixingcondition}, the projected gradients for the left modulation involve the combination
\(G_L-G_M\), whereas those for the right modulation involve \(G_R-G_M\).
 
 Explicitly, one obtains the Fourier-projected gradients in the form of discretized integrals, 
\[
G^{\rm Fourier}_{\delta_{ \alpha 0}} 
=
\Delta t
\sum_{j=1}^{N_{t}}
\rho_0
\left[
G_{\alpha}(t_j)-G_M(t_j)
\right],
\]
\[
G^{\rm Fourier}_{c_{\alpha \, n}}  = \Delta t
\sum_{j=1}^{N_{t}}
\rho_0
\left[
G_{\alpha}(t_j)-G_M(t_j)
\right]
\cos \left(\frac {2 \pi n t_j} {T} \right) ,
\]
and
\[
G^{\rm Fourier}_{d_{\alpha \, n}} 
= \Delta t
\sum_{j=1}^{N_{t}}
\rho_0
\left[
G_{\alpha}(t_j)-G_M(t_j)
\right]
\sin \left(\frac {2 \pi n t_j} {T} \right) \,. 
\]
for \(\alpha=L,R\) and \(1\leq n\leq N_{\rm modes}\). These projected gradients lead to the coefficient update at each iteration,
\begin{eqnarray}
\delta_{\alpha \, 0}^{(k+1)}
& = &
\delta_{\alpha \, 0}^{(k)}
+
\eta\,G^{\rm Fourier}_{\delta_{ \alpha 0}} \nonumber \\
c_{\alpha \, n}^{(k+1)}
& = &
c_{\alpha \, n}^{(k)}
+
\eta\,G^{\rm Fourier}_{c_{ \alpha n}}   \nonumber \\
d_{\alpha \, n}^{(k+1)}
& = &
d_{\alpha \, n}^{(k)}
+
\eta\,G^{\rm Fourier}_{d_{ \alpha n}}  
\end{eqnarray}
The gradients $G_{\alpha}(t_j)$, with $\alpha=L,M,R$, are obtained  from  Eq.~\eqref{eq:gradientobtention} using the forward propagation of the state [Eq.\eqref{eq:statepropagation}] together with the backward propagation of the adjoint state [Eqs.(\ref{eq:adjoint_terminal},\ref{eq:adjoint_propagation})]. The step size \(\eta\)
is chosen by the same line-search procedure as in the direct time-bin
implementation. The number of grid points $N_t$ no longer determines the dimension of the control space, which is set instead by the number of retained Fourier modes. One can therefore use a finer time grid than in a typical direct PMP optimization.

\bibliographystyle{quantum}
\bibliography{Topological_Transfer_Ref}

\end{document}